\documentclass[preprint,authoryear,12pt]{elsarticle}

\usepackage{amssymb}
\usepackage{amssymb}
\usepackage{latexsym}
\usepackage{amsmath}
\usepackage{graphicx}

\begin{document}

\begin{frontmatter}



\title{Mixed-variational formulation for phononic band-structure calculation of arbitrary unit cells}


\author{Ankit Srivastava}
\address{Department of Mechanical, Materials, and Aerospace Engineering,\\
Illinois Institute of Technology, Chicago, IL, 60616 USA}

\author{Sia Nemat-Nasser}
\address{Department of Mechanical and Aerospace Engineering\\ 
University of California, San Diego\\
La Jolla, CA, 92093-0416 USA\\}%

\begin{abstract}

This paper presents phononic band-structure calculation results obtained using a mixed variational formulation for 1-, and 2-dimensional unit cells. The formulation itself is presented in a form which is equally applicable to 3-dimensiomal cases. It has been established that the mixed-variational formulation presented in this paper shows faster convergence with considerably greater accuracy than variational principles based purely on the displacement field, especially for problems involving unit cells with discontinuous  constituent properties. However, the application of this formulation has been limited to fairly simple unit cells. In this paper we have extended the scope of the formulation by employing numerical integration techniques making it applicable for the evaluation of the phononic band-structure of unit cells displaying arbitrary complexity in their Bravais structure and in the shape, size, number, and anisotropicity of their micro-constituents. The approach is demonstrated through specific examples.


\end{abstract}

\end{frontmatter}




\section{Introduction}

There has been a recent surge of research effort towards achieving exotic dynamic response through novel  microstructural  design of composites. Within mechanics and elastodynamics these responses can be categorized in two broad areas: phononics and metamaterials. Phononics is the study of stress wave propagation in periodic elastic composites, whereas, metamaterials builds upon the area of phononics with dynamic homogenization schemes and seeks to create periodic composites with overall dynamic properties that are not shared by common materials. The required first step to attain this is to evaluate the phononic band-structure of periodic composites.

The phononic band-structure (\cite{martinezsala1995sound}) results from the periodic modulation of stress waves, and as such has deep similarities with areas like electronic band theory (\cite{bloch1928quantum}) and photonics (\cite{ho1990existence}). Such periodic modulations provide for very rich wave-physics and the potential for novel applications (\cite{cervera2001refractive,yang2002ultrasound,khelif2003trapping,reed2003reversed,yang2004focusing,gorishnyy2005hypersonic,mohammadi2008evidence,sukhovich2008negative,lin2009gradient}). These applications depend upon the ability of calculating the required phononic band-structure. In addition to the  ability of calculating phononic band-structures, certain research areas such as phononic band-structure optimization (\cite{sigmund2003systematic,rupp2007design,bilal2011ultrawide,diaz2005design,halkjaer2006maximizing}) and inverse problems in dynamic homogenization, also demand that the band-structure calculating algorithm possess speed, efficiency, accuracy, and versatility. There exist several techniques by which band-structures of photonic and phononic composites can be computed. These include the plane wave expansion (PWE) method (\cite{ho1990existence,leung1990full,zhang1990electromagnetic}), the multiple scattering method (\cite{kafesaki1999multiple}), the finite difference time domain method (\cite{chan1995order}), the finite element method (\cite{white1989finite}), variational methods (\cite{goffaux2003two}) and more (See \cite{hussein2009reduced}). 

In this paper we elaborate upon a mixed variational formulation for phononic band-structure calculations which is based upon varying both the displacement and the stress fields
(\cite{nemat1972harmonic}; \cite{nemat1975harmonic,minagawa1976harmonic,nemat2011homogenization}). 
Since it is based on a variational principle, any set of approximating functions can be used for calculations, e.g., plane-waves Fourier series or finite elements (\cite{minagawa1981finite}). The mixe formulation yields very accurate results and the rate of convergence of the corresponding approximating series solution is greater than that of the Rayleigh quotient with displacement-based approximating functions (\cite{babuska1978numerical}). Although the mixed-formulation shows a fast convergence, it has not yet been used to evaluate the band-structures of complex 2-, and 3-dimensional unit cells. In this paper we extend the scope of the formulation by employing numerical integrations and describe clearly how it can be applied to 1-, 2-, and 3-dimensional unit cells of arbitrary complexity in their Bravais structure and in the shape, size, number, and anisotropicity of their micro-constituents. We present 1-, and 2-dimensional test cases which verify the results of the formulation with published results in literature (exact solution for 1-dimensional and plane wave approximation for 2-dimensional). For the 2-phase 2-dimensional case we note that acceptable convergence over the first 18 phononic branches is achieved when the displacement and stress fields are approximated by 121 Fourier terms each. 

\section{Statement of the problem}

\begin{figure}[htp]
\centering
\includegraphics[scale=.6]{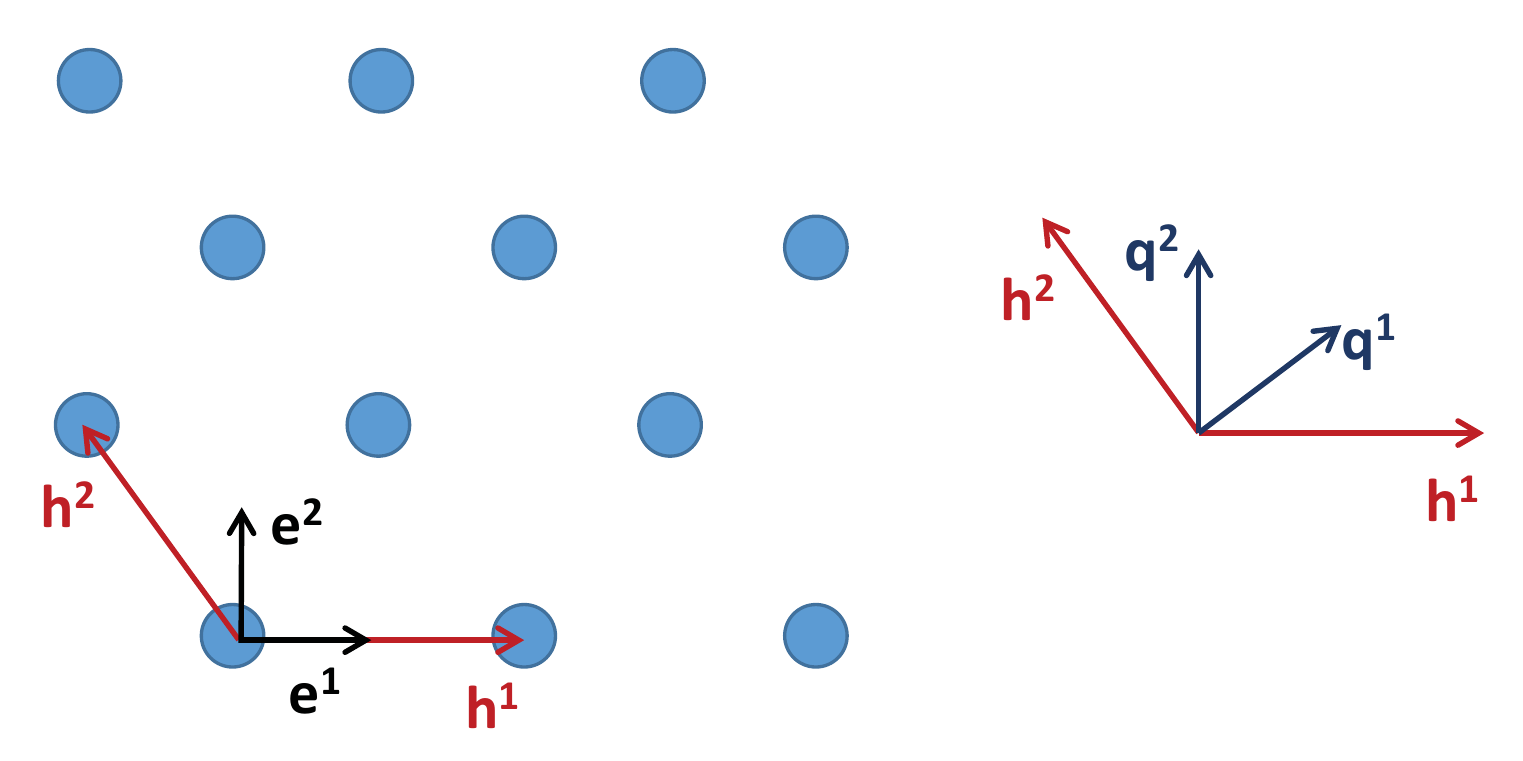}
\caption{Schematic of a 2-dimensional periodic composite. The unit cell vectors ($\mathbf{h}^1,\mathbf{h}^2$), reciprocal basis vectors ($\mathbf{q}^1,\mathbf{q}^2$), and the orthogonal vectors ($\mathbf{e}^1,\mathbf{e}^2$) are shown.}\label{fVectors}
\end{figure}

In the following treatment repeated Latin indices mean summation, whereas, repeated Greek indices do not. Consider the problem of elastic wave propagation in a general 3-dimensional periodic composite. The unit cell of the periodic composite is denoted by $\Omega$ and is characterized by 3 base vectors $\mathbf{h}^i$, $i=1,2,3$. Any point within the unit cell can be uniquely specified by the vector $\mathbf{x}=H_i\mathbf{h}^i$ where $0\leq H_i\leq 1, i=1,2,3$. The same point can also be specified in the orthogonal basis as $\mathbf{x}=x_i\mathbf{e}^i$. The reciprocal base vectors of the unit cell are given by,
\begin{equation}
\mathbf{q}^1=2\pi\frac{\mathbf{h}^2\times\mathbf{h}^3}{\mathbf{h}^1\cdot(\mathbf{h}^2\times\mathbf{h}^3)};\quad \mathbf{q}^2=2\pi\frac{\mathbf{h}^3\times\mathbf{h}^1}{\mathbf{h}^2\cdot(\mathbf{h}^3\times\mathbf{h}^1)};\quad \mathbf{q}^3=2\pi\frac{\mathbf{h}^1\times\mathbf{h}^2}{\mathbf{h}^3\cdot(\mathbf{h}^1\times\mathbf{h}^2)}
\end{equation}
such that $\mathbf{q}^i\cdot\mathbf{h}^j=2\pi\delta_{ij}$, where the denominators of the above vectors are the volume of the unit cell. Fig. (\ref{fVectors}) is the schematic of a 2-dimensional unit cell, indicating the unit cell basis vectors, the reciprocal basis vectors and the orthogonal basis vectors.

The wave vector for a Bloch-wave traveling in the composite are given as $\mathbf{k}=Q_i\mathbf{q}^i$ where $0\leq Q_i\leq 1,i=1,2,3$. The composite is characterized by a spatially varying stiffness tensor, $C_{jkmn}(\mathbf{x})$, and density, $\rho(\mathbf{x})$, which satisfy the following periodicity conditions:
\begin{equation}
C_{jkmn}(\mathbf{x}+n_i\mathbf{h}^i)=C_{jkmn}(\mathbf{x});\quad \rho(\mathbf{x}+n_i\mathbf{h}^i)=\rho(\mathbf{x}),
\end{equation}
where $n_i$$(i=1,2,3)$ are integers.

\subsection{Field equations and boundary conditions}

For harmonic elastodynamic problems the equations of motion and kinematic relations at any point $\mathbf{x}$ in $\Omega$ are given by
\begin{equation}\label{equationofmotion}
\sigma_{jk,k}=-\lambda\rho u_j; \quad \varepsilon_{jk}=\frac{1}{2}(u_{j,k}+u_{k,j}),
\end{equation}
where $\lambda=\omega^2$, and $\boldsymbol{\sigma}e^{-i\omega t},\boldsymbol{\varepsilon}e^{-i\omega t}$, and $ \mathbf{u}e^{-i\omega t}$ are the space and time dependent stress tensor, strain tensor, and displacement vector, respectively. The stress tensor is related to the strain tensor through the elasticity tensor, $\sigma_{jk}=C_{jkmn}\varepsilon_{mn}$. The traction and displacement at any point in the composite are related to the corresponding traction and displacement at another point, sperated from the first by a unit cell, through Bloch relations. These relations serve as the homogeneous boundary conditions on $\partial\Omega$. If the Bloch wave vector is $\mathbf{k}$ then these boundary conditions are given by,
\begin{equation}\label{boundaryconditions}
u_j(\mathbf{x}+\mathbf{h}^i)=u_j(\mathbf{x})e^{i\mathbf{k}\cdot\mathbf{h}^i};\quad t_j(\mathbf{x}+\mathbf{h}^i)=-t_j(\mathbf{x})e^{i\mathbf{k}\cdot\mathbf{h}^i}, \quad \mathbf{x}\in\partial\Omega,
\end{equation}
where $t_j=\sigma_{jk}\nu_k$ are the components of the traction vector and $\boldsymbol{\nu}$ is the exterior unit normal vector on $\partial\Omega$.

\subsection{Mixed-variational formulation}

It has been shown (\cite{nemat1975harmonic}, \cite{minagawa1976harmonic}) that the solution to (\ref{equationofmotion}) that satisfies the boundary conditions, (\ref{boundaryconditions}), renders the following functional stationary:
\begin{equation}\label{mixedvariational}
\lambda_N=\frac{\langle\sigma_{jk},u_{j,k}\rangle+\langle u_{j,k},\sigma_{jk}\rangle+\langle D_{jkmn}\sigma_{jk},\sigma_{mn}\rangle}{\langle\rho u_j,u_j\rangle},
\end{equation}
where $\mathbf{D}$ is the compliance tensor and the inner product is given by,
\begin{equation}
\langle u,v\rangle=\int_\Omega uv^*d\Omega,
\end{equation}
where $v^*$ is the complex conjugate of $v$.

\subsection{Approximation with periodic test functions}

Now we approximate the stress and displacement fields with the following test functions:
\begin{equation}\label{approximation}
\bar{u}_j=\sum_{\alpha,\beta,\gamma}U^{\alpha\beta\gamma}_jf^{\alpha\beta\gamma}(\mathbf{x}),\quad \bar{\sigma}_{jk}=\sum_{\alpha,\beta,\gamma}S^{\alpha\beta\gamma}_{jk}f^{\alpha\beta\gamma}(\mathbf{x}),
\end{equation}
where the test functions satisfy the boundary conditions, (\ref{boundaryconditions}), and are orthogonal in the sense that $\langle f^{\alpha\beta\gamma},f^{\theta\eta\xi}\rangle$ is proportional to $\delta_{\alpha\theta}\delta_{\beta\eta}\delta_{\gamma\xi}$, $\boldsymbol{\delta}$ being the Kronecker delta. Substituting from (\ref{approximation}) to (\ref{mixedvariational}) and setting the derivative of $\lambda_N$ with respect to the unknown coefficients, ($U^{\alpha\beta\gamma}_j,S^{\alpha\beta\gamma}_{jk}$), equal to zero, we arrive at the following system of linear homogeneous equations:
\begin{align}\label{equationshomogeneous}
\langle\bar{\sigma}_{jk,k}+\lambda_N\rho\bar{u}_j,f^{\theta\eta\xi}\rangle=0\nonumber,\\
\langle D_{jkmn}\bar{\sigma}_{mn}-\bar{u}_{(j,k)},f^{\theta\eta\xi}\rangle=0\nonumber,\\
j,k,m,n=1,2,3,
\end{align}
where $\bar{u}_{(j,k)}\equiv\bar{\varepsilon}_{jk}=\frac{1}{2}(\bar{u}_{j,k}+\bar{u}_{k,j})$. For the general 3-dimensional case, if $\alpha,\beta,\gamma,\theta,\eta,\xi$ vary from $-M$ to $M$ then (\ref{equationshomogeneous}) represents $9(2M+1)^3$ linear homogeneous equations in the $9(2M+1)^3$ unknown displacement and stress coefficients. Given the symmetry of the stress tensor, these coefficients are $3(2M+1)^3$ number of $U^{\alpha\beta\gamma}_j$ and $6(2M+1)^3$ number of independent $S^{\alpha\beta\gamma}_{jk}$.

To approximate the stress and displacement fields in (\ref{approximation}), we use test functions of the following form:
\begin{equation}
f^{\alpha\beta\gamma}(\mathbf{x})=e^{i(\mathbf{k}\cdot\mathbf{x}+2\pi[\alpha H_1+\beta H_2+\gamma H_3])},
\end{equation}
where $\mathbf{x}=H_j\mathbf{h}^j$.

\subsection{Suitability of the test functions}

In order for the test functions to be suitable they should satisfy the boundary conditions, (\ref{boundaryconditions}), and should be orthogonal in the sense mentioned above. To show that these test functions satisfy the boundary conditions we note for $\mathbf{x}'=\mathbf{x}+\mathbf{h}^k=(H_j+\delta_{jk})\mathbf{h}^j$ that
\begin{equation}
f^{\alpha\beta\gamma}(\mathbf{x}')=e^{i(\mathbf{k}\cdot\mathbf{x}+2\pi[\alpha H_1+\beta H_2+\gamma H_3])}e^{i(\mathbf{k}\cdot(\mathbf{h}^j\delta_{jk}))}e^{i(2\pi[\alpha\delta_{1k}+\beta\delta_{2k}+\gamma\delta_{3k}])}.
\end{equation}
Since $\alpha,\beta,\gamma$ are integers the last term equals unity and we have
\begin{equation}
f^{\alpha\beta\gamma}(\mathbf{x}+\mathbf{h}^k)=f^{\alpha\beta\gamma}(\mathbf{x})e^{i(\mathbf{k}\cdot\mathbf{h}^k)},
\end{equation}
showing that test functions satisfy the boundary conditions. To show orthogonality we note that
\begin{align}
\nonumber \langle f^{\alpha\beta\gamma},f^{\theta\eta\xi}\rangle=\int_\Omega e^{i(\mathbf{k}\cdot\mathbf{x}+2\pi[\alpha H_1+\beta H_2+\gamma H_3])} e^{-i(\mathbf{k}\cdot\mathbf{x}+2\pi[\theta H_1+\eta H_2+\xi H_3])}d\Omega\\
=\int_\Omega e^{i(2\pi[(\alpha-\theta) H_1+(\beta-\eta) H_2+(\gamma-\xi) H_3])}d\Omega\propto\delta_{\alpha\theta}\delta_{\beta\eta}\delta_{\gamma\xi},
\end{align}
showing the orthogonality of the test functions.

\subsection{Derivatives of the test functions}

A special note of consideration here regards the spatial derivatives of the test function. The set of linear homogeneous equations (\ref{equationshomogeneous}) involve derivatives of the test functions in the orthogonal coordinate system. It is, therefore, necessary to express the test functions in these coordinates. To do this we first express $\mathbf{k}$ in the reciprocal basis ($\mathbf{k}=Q_i\mathbf{q}^i$) and $\mathbf{x}$ in the unit cell basis ($\mathbf{x}=H_j\mathbf{h}^j$) and by using $\mathbf{q}^i\cdot\mathbf{h}^j=2\pi\delta_{ij}$ we note that,
\begin{equation}
f^{\alpha\beta\gamma}(\mathbf{x})=e^{i2\pi[(Q_1+\alpha) H_1+(Q_2+\beta) H_2+(Q_3+\gamma) H_3]}.
\end{equation}
Now $H_1,H_2,H_3$ are expressed in the orthogonal coordinate system. Let $\mathbf{x}=H_j\mathbf{h}^j=x_k\mathbf{e}^k$ where $\mathbf{e}^k$ is the $k^{\mathrm{th}}$ unit vector in the orthogonal coordinate system and $x_k$ is the $k^{\mathrm{th}}$ coordinate of the vector $\mathbf{x}$ in this system. By taking a dot product of this equation with the three orthogonal unit vectors we can express $H_j$ in terms of $x_k$,
\begin{equation}
\{H_1 H_2 H_3\}^\mathrm{T}=[A
]^{-1}\{x_1 x_2 x_3\}^\mathrm{T}\equiv [T
]\{x_1 x_2 x_3\}^\mathrm{T},
\end{equation}
where the $3\times 3$ matrix $[A]$ has components $A_{jk}=\mathbf{h}^j\cdot\mathbf{e}^k$. Now the test function can be written as,
\begin{equation}
f^{\alpha\beta\gamma}(\mathbf{x})=e^{i2\pi Q^{\alpha\beta\gamma}_k x_k},
\end{equation}
where
\begin{align}
\nonumber Q^{\alpha\beta\gamma}_1=T_{11}(Q_1+\alpha)+T_{12}(Q_2+\beta)+T_{13}(Q_3+\gamma),\\
\nonumber Q^{\alpha\beta\gamma}_2=T_{21}(Q_1+\alpha)+T_{22}(Q_2+\beta)+T_{23}(Q_3+\gamma),\\
Q^{\alpha\beta\gamma}_3=T_{31}(Q_1+\alpha)+T_{32}(Q_2+\beta)+T_{33}(Q_3+\gamma).
\end{align}
The $j^\mathrm{th}$ derivative of the test function is now given by:
\begin{equation}
f^{\alpha\beta\gamma}_{,j}=(i2\pi Q^{\alpha\beta\gamma}_{k}\delta_{kj})f^{\alpha\beta\gamma}.
\end{equation}

\section{Numerical solution}

The band-structure of the composite is given by the $\mathbf{q}-\omega$ pairs which lead to nontrivial solutions of (\ref{equationshomogeneous}). To calculate these pairs, (\ref{equationshomogeneous}) is first written in the following equivalent matrix form:
\begin{align}\label{equationshomogeneousMatrix}
\nonumber \mathbf{HS}+\lambda_N\mathbf{\Omega U}=0,\\
\mathbf{\Phi S}+\mathbf{H^*U}=0.
\end{align}
Column vectors $\mathbf{S},\mathbf{U}$ contain the unknown coefficients of the periodic expansions of stress and displacement, respectively. Matrices $\mathbf{H},\mathbf{\Omega},\mathbf{\Phi},\mathbf{H}^*$ contain the integrals of the various functions appearing in (\ref{equationshomogeneous}). Their sizes depend upon whether the problem under consideration is 1-, 2-, or 3-dimensional. These matrices would be described more clearly in the subsequent sections in which numerical examples are shown. The above system of equations can be recast into the following traditional eigenvalue problem:
\begin{equation}\label{eigenvalueproblem}
(\mathbf{H}\mathbf{\Phi}^{-1}\mathbf{H}^*)^{-1}\mathbf{\Omega}\mathbf{U}=\frac{1}{\lambda_N}\mathbf{U},
\end{equation}
whose eigenvalue solutions represent the frequencies ($\omega_N=\sqrt{\lambda_N}$) associated with the wave-vector under consideration ($\mathbf{q}$). The eigenvectors of the above equation are used to calculate the displacement modeshapes from (\ref{approximation}). The relation $\mathbf{S}=-\mathbf{\Phi}^{-1}\mathbf{H}^*\mathbf{U}$ is used to evaluate the stress eigenvector which is subsequently used to calculate the stress modeshape from (\ref{approximation}).

The integrals occurring in (\ref{equationshomogeneousMatrix}) are numerically calculated over $\Omega$. This is a diversion from the earlier results (\cite{nemat1975harmonic},  \cite{minagawa1976harmonic}) where the closed form solutions of the integrals were used. By numerically evaluating the integrals we have extended the scope of the method to include arbitrary inclusions and all possible Bravais lattices in 1-, 2-, and 3-dimensions. Numerical integration is achieved by dividing the domain $\Omega$ into $P$ subdomains $\Omega_i,i=1,2...P$. The volume integral of any function $F(\mathbf{x})$ is then approximated as,
\begin{equation}
\int_\Omega F(\mathbf{x})d\Omega=\sum_i^PF_iV_i,
\end{equation}
where $F_i$ is the value of the function $F(\mathbf{x})$ evaluated at the centroid of $\Omega_i$ and $V_i$ is the volume of $\Omega_i$. For meshing in 2-, and 3-D we have used a freely available Finite Element software \cite{geuzaine2009gmsh} which automatically discretizes the domain in triangular (for 2-D) and tetrahedral (in 3-D) elements and provides the nodal position and elemental connectivity matrices. MATLAB (or Python) routines are then used to calculate the volumes and centroids of these subdomains and the required integrals over $\Omega$.

\section{Application to 1-D periodic composites}

There is only one possible Bravais lattice in 1-dimension with a unit cell vector whose length equals the length of the unit cell itself. Without any loss of generality we take the direction of this vector to be the same as $\mathbf{e}^1$. If the length of the unit cell is $a$, then we have $\mathbf{h}^1=a\mathbf{e}^1$. The reciprocal vector is given by $\mathbf{q}^1=(2\pi/a)\mathbf{e}^1$. The wave-vector of a Bloch wave traveling in this composite is specified as $\mathbf{k}=Q_1\mathbf{q}^1$. To completely characterize the band-structure of the unit cell it is sufficient to evaluate the dispersion relation in the irreducible Brillouin zone ($-.5\leq Q_1\leq .5$).

For plane longitudinal wave propagating in the $\mathbf{e}^1$ direction the only displacement component of interest is $u_1$ and the only relevant stress component is $\sigma_{11}$ (for plane shear waves traveling in $\mathbf{e}^1$ direction the quantities of interest are $u_2$ and $\sigma_{12}$, and for the anti-plane shear waves, they are $u_3$ and $\sigma_{13}$, respectively). The equation of motion and the constitutive relation are,
\begin{equation}\label{equationofmotion1D}
\sigma_{11,1}=-\lambda\rho(x_1) u_1, \quad \sigma_{11}=E(x_1)u_{1,1},
\end{equation}
where $E(x_1)$ is the spatially varying Young's modulus.

These are approximated by 1-D periodic functions,
\begin{equation}\label{approximation1d}
\bar{u}_1=\sum_{\alpha=-M}^MU^{\alpha}_1e^{i(\mathbf{k}\cdot\mathbf{x}+2\pi\alpha H_1)},\quad \bar{\sigma}_{11}=\sum_{\alpha=-M}^MS^{\alpha}_{11}e^{i(\mathbf{k}\cdot\mathbf{x}+2\pi\alpha H_1)},
\end{equation}
which can be further simplified to,
\begin{equation}\label{approximation1dS}
\bar{u}_1=\sum_{\alpha=-M}^MU^{\alpha}_1e^{i2\pi(Q_1+\alpha)x_1/a},\quad \bar{\sigma}_{11}=\sum_{\alpha=-M}^MS^{\alpha}_{11}e^{i2\pi(Q_1+\alpha)x_1/a}.
\end{equation}
Now the system of equations denoting the eigenvalue problem are written as:
\begin{align}\label{equationshomogeneous1D}
\langle\bar{\sigma}_{11,1}+\lambda_N\rho\bar{u}_1,f^{\theta}\rangle=0\nonumber,\\
\langle D\bar{\sigma}_{11}-\bar{u}_{1,1},f^{\theta}\rangle=0,
\end{align}
where $M\leq\alpha,\theta\leq M$ and $D=1/E$. The above are transformed to the matrix form of (\ref{equationshomogeneousMatrix}) with the following column vectors:
\begin{align}
\mathbf{U}=\{U^{-M}_1\;...\;U^{0}_1\;...\;U^{M}_1\}^T\nonumber,\\
\mathbf{S}=\{S^{-M}_{11}\;...\;S^{0}_{11}\;...\;S^{M}_{11}\}^T.
\end{align}
The associated coefficient matrices have the following elements:
\begin{align}\label{coefficients1d}
\nonumber[\mathbf{H}]_{mn}=\frac{i2\pi(Q_1+m-M-1)}{a}\int_0^ae^{i2\pi(m-n)x_1/a}dx_1,\\
\nonumber[\mathbf{\Omega}]_{mn}=\int_0^a\rho(x_1)e^{i2\pi(m-n)x_1/a}dx_1,\\
\nonumber[\mathbf{\Phi}]_{mn}=\int_0^aD(x_1)e^{i2\pi(m-n)x_1/a}dx_1,\\
\nonumber[\mathbf{H}]^*_{mn}=\frac{-i2\pi(Q_1+m-M-1)}{a}\int_0^ae^{i2\pi(m-n)x_1/a}dx_1,\\
m,n=1,2,...(2M+1).
\end{align}
It must be noted that given the periodicity of the exponential function, matrices $[\mathbf{H}] $ and $[\mathbf{H}]^*$ are diagonal. Now for given values of $Q_1$, the eigenvalue problem can be solved for the frequencies $\omega_N$.

\subsection{2-phase layered composite: comparison with Rytov solution}

The problem of wave propagation in 1-dimension layered composite can be solved exactly. For a 2-phase composite this solution was first given by Rytov \cite{rytov1956acoustical}. Here we show a comparison of the results from the mixed-variational formulation with the exact Rytov solution. Since for the 1-dimensional case the integrals in (\ref{coefficients1d}) can be calculated exactly, we have not employed numerical integration via discretization of $\Omega$.


\begin{figure}[htp]
\centering
\includegraphics[scale=.65]{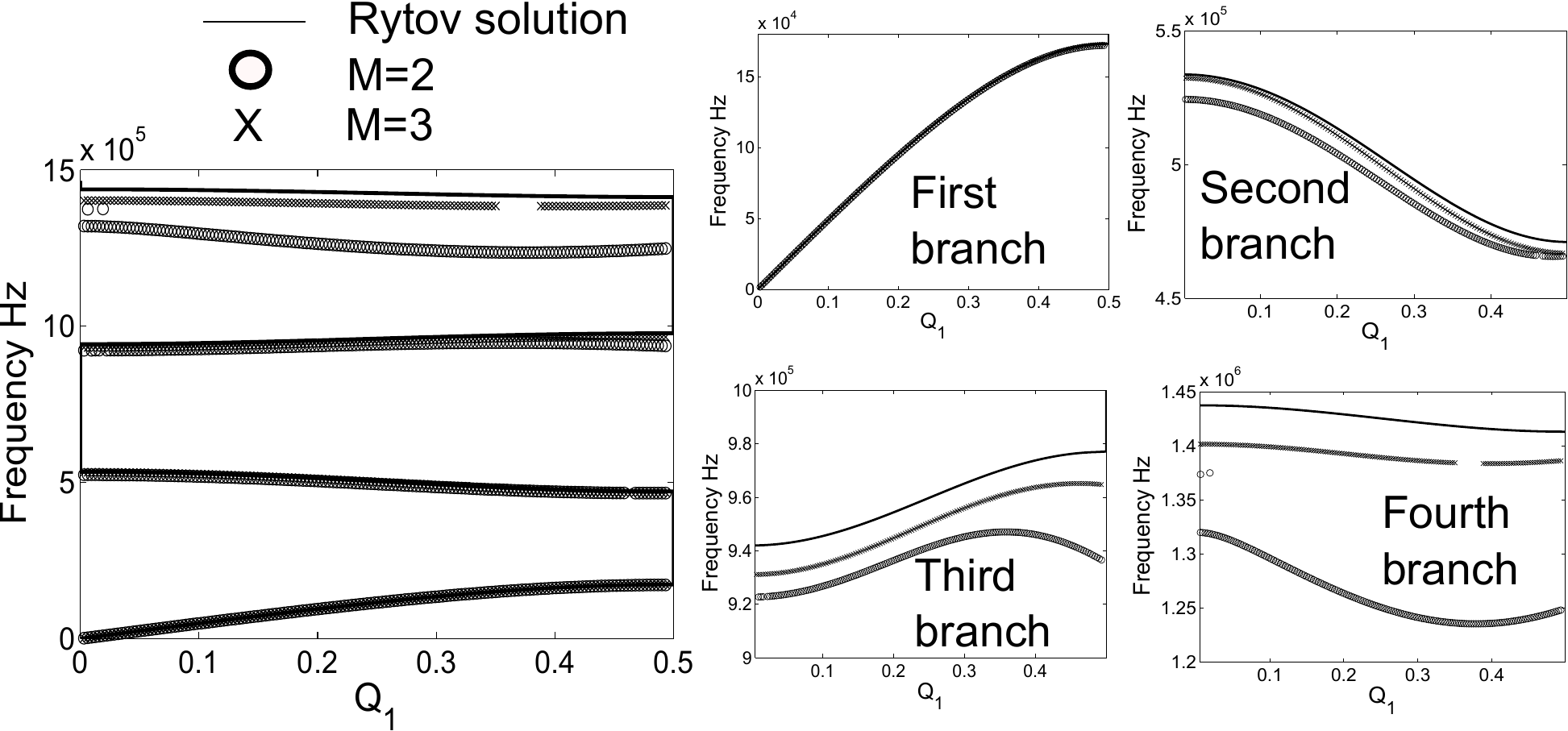}
\caption{Comparison of the mixed-variational formulation results with Rytov solution. The first 4 branches are shown for $M=2,3$.}\label{fCompRytov}
\end{figure}

The composite under consideration is a 2-phase composite with $\Omega$ consisting of the following 2 phases:
\begin{enumerate}
\item Phase 1: $E=8$ Gpa, $\rho=1000$ kg/m$^3$, thickness=0.003 m
\item Phase 2: $E=300$ Gpa, $\rho=8000$ kg/m$^3$, thickness=0.001 m
\end{enumerate}
Fig. (\ref{fCompRytov}) shows the comparison of the results from the mixed-variational formulation with the Rytov solution for the first four branches of the composite. The results are calculated for $0\leq Q_1\leq.5$ at a total of 155 points. The mixed-variational results are shown for $M=2,3$. At each value of $Q_1$ the eigenvalue solution results in $2M+1$ frequencies. This process is repeated for the 155 values of $Q_1$  between 0 and 0.5. From the results in Fig. (\ref{fCompRytov}) it is clear that the first two branches are captured well even with $M=2$ whereas higher branches require more terms in the periodic expansion for adequate accuracy. Convergence to the exact solution is achieved  for all branches as higher number of terms are used in the expansion.


\begin{figure}[htp]
\centering
\includegraphics[scale=.4]{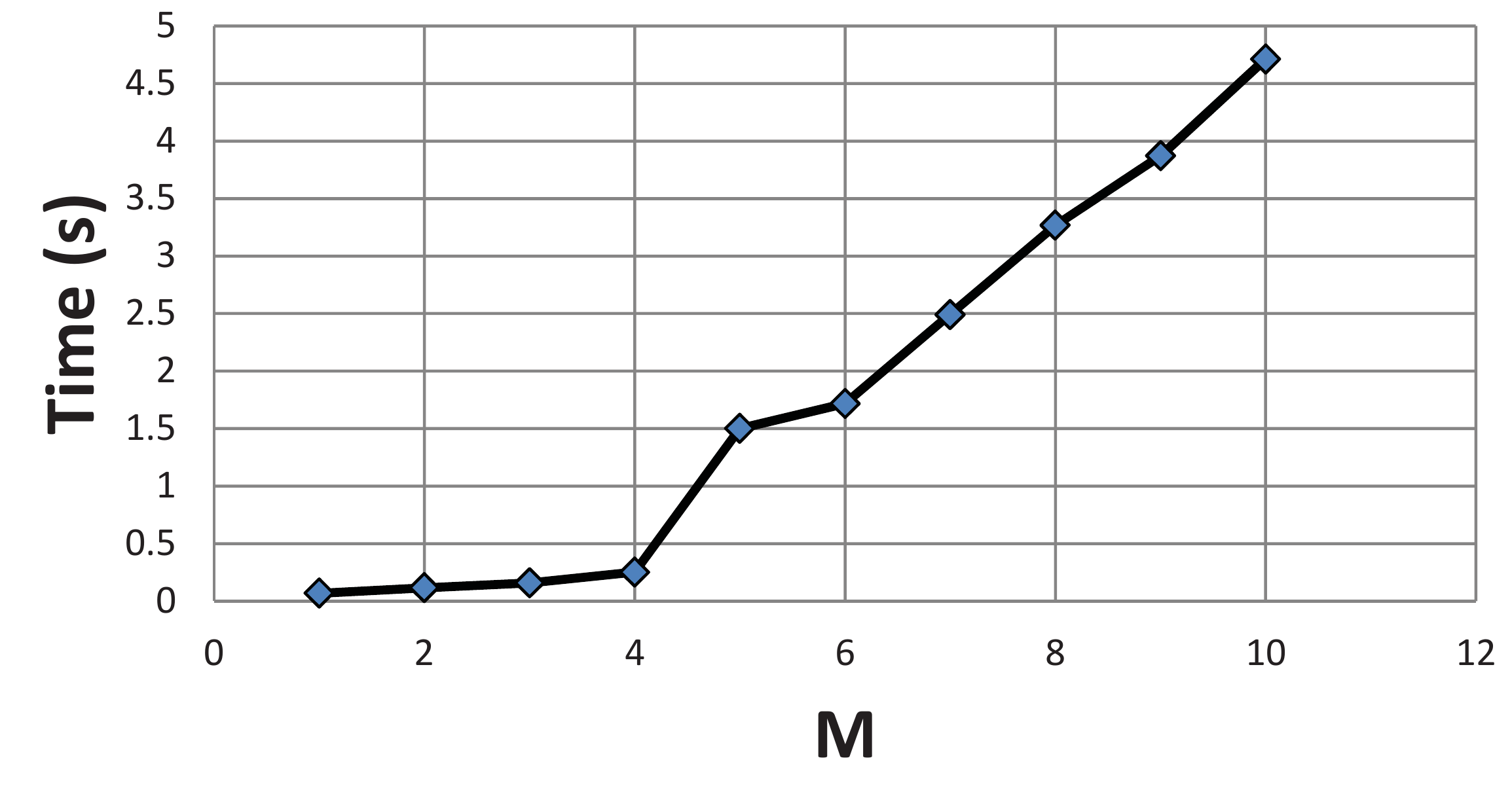}
\caption{Computation time for $M=1,2,...10$}\label{fCompTime1D}
\end{figure}

The cpu-time as calculated from the cputime command in MATLAB gives an average value of 0.11544 seconds for $M=2$ computations and 0.15912 seconds for $M=3$ computations. It must be stressed that in each case this cpu-time refers to the time it takes to calculate all $2M+1$ frequencies for all 155 $Q_1$-points. Fig. (\ref{fCompTime1D}) shows the average computation time as given by the cpu-time command for $M$ ranging from 1 to 10.

\section{Application to 2-D periodic composites}

There are five possible Bravais lattices in 2 dimensions. However, they can be specified using two unit cell vectors ($\mathbf{h}^1,\mathbf{h}^2$). The reciprocal vector are $\mathbf{q}^1,\mathbf{q}^2$. The wave-vector of a Bloch-wave traveling in this composite is specified as $\mathbf{k}=Q_1\mathbf{q}^1+Q_2\mathbf{q}^2$. To characterize the band-structure of the unit cell we evaluate the dispersion relation along the boundaries of the irreducible Brillouin zone ($0\leq Q_1\leq .5,Q_2=0;\;Q_1=.5,0\leq Q_2\leq .5;\;0\leq Q_1\leq .5,Q_2=Q_1$). In traditional notation these boundaries are specified as $\Gamma-X,X-M,M-\Gamma$ respectively. Therefore, we would be considering band-structure over $\Gamma-X-M-\Gamma$.

For purposes of demonstration and comparison we consider the case of plane-strain state in the composite. The relevant stress components for the plane-strain case are $\sigma_{11},\sigma_{22},\sigma_{12}$ and the relevant displacement components are $u_1,u_2$. The equations of motion and the constitutive relation are,
\begin{equation}\label{equationofmotion2D}
\sigma_{jk,k}=-\lambda\rho(\mathbf{x}) u_j, \quad D_{jkmn}(\mathbf{x})\sigma_{mn}=u_{j,k},\quad j,k,m,n=1,2,
\end{equation}
where $\mathbf{D}$ is the compliance tensor. For an isotropic material in plane strain, $\mathbf{D}$ is given by,
\begin{equation}
D_{jkmn}=\frac{1}{2\mu}\left[\frac{1}{2}(\delta_{jm}\delta_{kn}+\delta_{jn}\delta_{km})-\frac{\lambda}{2(\mu+\lambda)}\delta_{jk}\delta_{mn}\right],\quad j,k,m,n=1,2,
\end{equation}
where $\lambda,\mu$ are the Lame$^{'}$ constants. The stresses and displacements are approximated by the following 2-D periodic functions:
\begin{equation}\label{approximation2d}
\bar{u}_j=\sum_{\alpha,\beta=-M}^MU^{\alpha\beta}_{j}e^{i2\pi Q^{\alpha\beta}_l x_l},\quad \bar{\sigma}_{jk}=\sum_{\alpha,\beta=-M}^MS^{\alpha\beta}_{jk}e^{i2\pi Q^{\alpha\beta}_l x_l};\quad j,k,l=1,2,
\end{equation}
where
\begin{align}
\nonumber Q^{\alpha\beta}_1=T_{11}(Q_1+\alpha)+T_{21}(Q_2+\beta),\\
Q^{\alpha\beta}_2=T_{12}(Q_1+\alpha)+T_{22}(Q_2+\beta),
\end{align}
and the square matrix $[\mathbf{T}]$ is the inverse of the matrix $[\mathbf{A}]$ with components $[\mathbf{A}]_{jk}=\mathbf{h}^j\cdot\mathbf{e}^k$.

\subsection{Details of the matrices}

The matrix form of the eigenvalue problem is given by (\ref{equationshomogeneousMatrix}) with the following column vectors:
\begin{align}
\mathbf{U}=\{U^{\alpha\beta}_1\;U^{\alpha\beta}_2\}^T\nonumber,\\
\mathbf{S}=\{S^{\alpha\beta}_{11}\;S^{\alpha\beta}_{22}\;S^{\alpha\beta}_{12}\}^T.
\end{align}
Since $\alpha,\beta,\theta,\eta$ vary from $-M$ to $M$, the length of the column vector $\mathbf{U}$ is $2(2M+1)^2$ and the length of $\mathbf{S}$ is $3(2M+1)^2$. Corresponding to these column vectors, the size of $\mathbf{H}$ is $3(2M+1)^2\times 2(2M+1)^2$, $\mathbf{\Omega}$ is $2(2M+1)^2\times 2(2M+1)^2$, and $\mathbf{\Phi}$ is $3(2M+1)^2\times 3(2M+1)^2$. To clarify the contents of the matrices $[\mathbf{H}],[\mathbf{\Omega}],[\mathbf{\Phi}]$ we introduce the following modified coordinates:
\begin{align*}
I_1=(\alpha+M)(2M+1)+(\beta+1+M),\\
J_1=(\theta+M)(2M+1)+(\eta+1+M),\\
I_2=I_1+(2M+1)^2,\quad J_2=J_1+(2M+1)^2,\\
I_3=I_2+(2M+1)^2,\quad J_3=J_2+(2M+1)^2.
\end{align*}
Components of the $\mathbf{H}$ matrix are given by,
\begin{align}\label{eh}
[\mathbf{H}]_{I_1J_1}=i2\pi Q^{\alpha\beta}_1\int_\Omega fd\Omega,\quad [\mathbf{H}]_{I_2J_2}=i2\pi Q^{\alpha\beta}_2\int_\Omega fd\Omega,\\
[\mathbf{H}]_{I_1J_3}=[\mathbf{H}]_{I_2J_2},\quad [\mathbf{H}]_{I_2J_3}=[\mathbf{H}]_{I_1J_1}.
\end{align}
We also have $[\mathbf{H}]^*=-[\mathbf{H}]^T$ where the superscript $T$ denotes a matrix transpose. Components of the $\mathbf{\Omega}$ matrix are given by:
\begin{equation}\label{eo}
[\mathbf{\Omega}]_{I_1J_1}=\int_\Omega
\rho(x_1,x_2)fd\Omega;\quad[\mathbf{\Omega}]_{I_2J_2}=[\mathbf{\Omega}]_{I_1J_1}
\end{equation}
The rest of the terms in the $\mathbf{\Omega}$ matrix being zero. The components of the $\mathbf{\Phi}$ matrix are given by,
\begin{align}\label{ep}
\nonumber[\mathbf{\Phi}]_{I_1J_1}=\int_\Omega D_{1111}(x_1,x_2)fd\Omega, \quad [\mathbf{\Phi}]_{I_1J_2}=\int_\Omega D_{1122}(x_1,x_2)fd\Omega,\\ 
[\mathbf{\Phi}]_{I_1J_3}=2\int_\Omega D_{1112}(x_1,x_2)fd\Omega,\quad
\nonumber[\mathbf{\Phi}]_{I_2J_1}=\int_\Omega D_{2211}(x_1,x_2)fd\Omega,\\ [\mathbf{\Phi}]_{I_2J_2}=\int_\Omega D_{2222}(x_1,x_2)fd\Omega,\quad [\mathbf{\Phi}]_{I_2J_3}=2\int_\Omega D_{2212}(x_1,x_2)fd\Omega,\\
[\mathbf{\Phi}]_{I_3J_1}=2\int_\Omega D_{1211}(x_1,x_2)fd\Omega,\quad [\mathbf{\Phi}]_{I_3J_2}=2\int_\Omega D_{1222}(x_1,x_2)fd\Omega,\\
[\mathbf{\Phi}]_{I_3J_3}=4\int_\Omega D_{1212}(x_1,x_2)fd\Omega,
\end{align}
where $f\equiv e^{i2\pi[(Q^{\alpha\beta}_1-Q^{\theta\eta}_1)x_1+(Q^{\alpha\beta}_2-Q^{\theta\eta}_2)x_2]}$. The above relations are  for the most general case. For isotropic plane-strain case several compliance components go to zero. This would result in several components of the $\mathbf{\Phi}$ matrix going to zero. Additionally, given the periodicity of the exponential function, the off diagonal terms in $[\mathbf{H}],[\mathbf{H}]^*$ are zero.

\subsection{Example: square unit cells}

All calculations shown for the 2-D cases in this and the subsequent sections require the calculation of various integrals in Eqs. (\ref{eh},\ref{eo},\ref{ep}) over the unit cell $\Omega$. In a divergence from earlier studies (\cite{nemat1975harmonic}, \cite{minagawa1976harmonic}) which used closed form expressions for these integrals, we have employed numerical integrals over automatically generated subdomains of $\Omega$. These subdomains are generated by a freely available FE software GMSH \cite{geuzaine2009gmsh}. Since we are performing numerical integrations with a view towards making the formulation more versatile, we begin with presenting some comparisons with the original results shown by \cite{nemat1975harmonic}. These original calculations were performed for fiber-reinforced square unit cells.

\begin{figure}[htp]
\centering
\includegraphics[scale=.3]{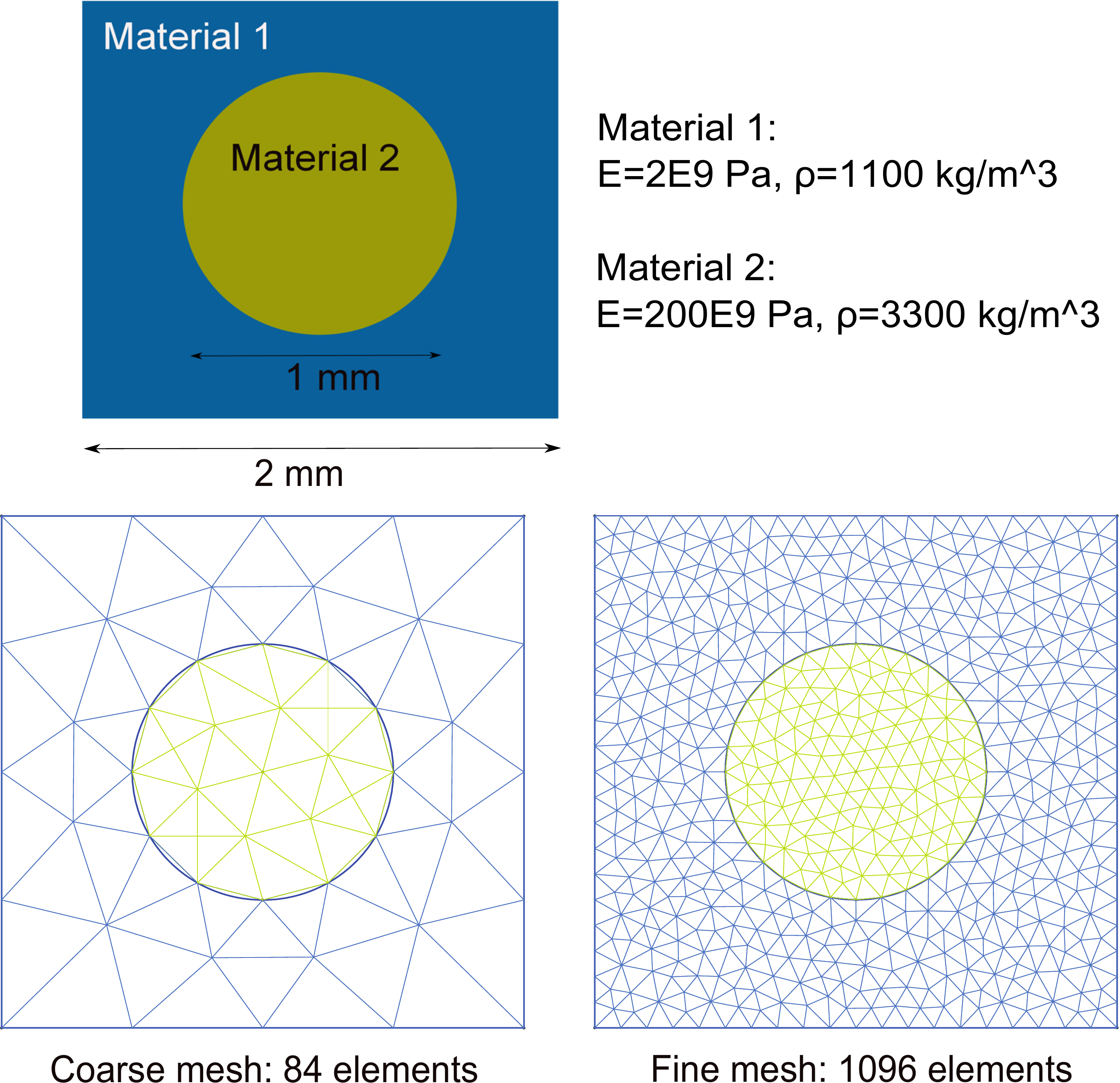}
\caption{Square unit cell with an embedded circular fiber.}\label{fUnitCellSquareCircFib}
\end{figure}

Consider, first, the case of a circular fiber embedded in a square unit cell of size 2 mm. The diameter of the fiber is half the length of the unit cell. The unit cell vectors are $\mathbf{h}^1=.002\mathbf{e}^1,\mathbf{h}^2=.002\mathbf{e}^2$. The materials making up the matrix and the fiber are taken as isotropic and the material properties are provided in Fig. (\ref{fUnitCellSquareCircFib}).

Note that with the dimensions and material properties provided in Fig. (\ref{fUnitCellSquareCircFib}), we satisfy the conditions used in (\cite{nemat1975harmonic}). These are $a_1/a_2=1,b_1/a_1=0.5,\theta=3,C_{1111}^{(2)}/C_{1111}^{(1)}=100$. Also, we have used a Poisson ratio of 0.3,  as in (\cite{nemat1975harmonic}). Fig. (\ref{fUnitCellSquareCircFib})  shows two levels of discretizations used for numerically evaluating the integrals. The unit cell is divided into 84 triangular elements at the level of the coarse dicretization and into 1096 elements at the finer discretization. In each case the integrals are calculated by evaluating the respective functions at the centroid of each triangle, multiplying it with the area of that triangle, and summing it over all the triangles which constitute the unit cell. In this way the matrices forming the eigenvalue problem are assembled. The eig command in MATLAB is finally used to solve for the eigenvalues of the system.


\begin{figure}[htp]
\centering
\includegraphics[scale=.45]{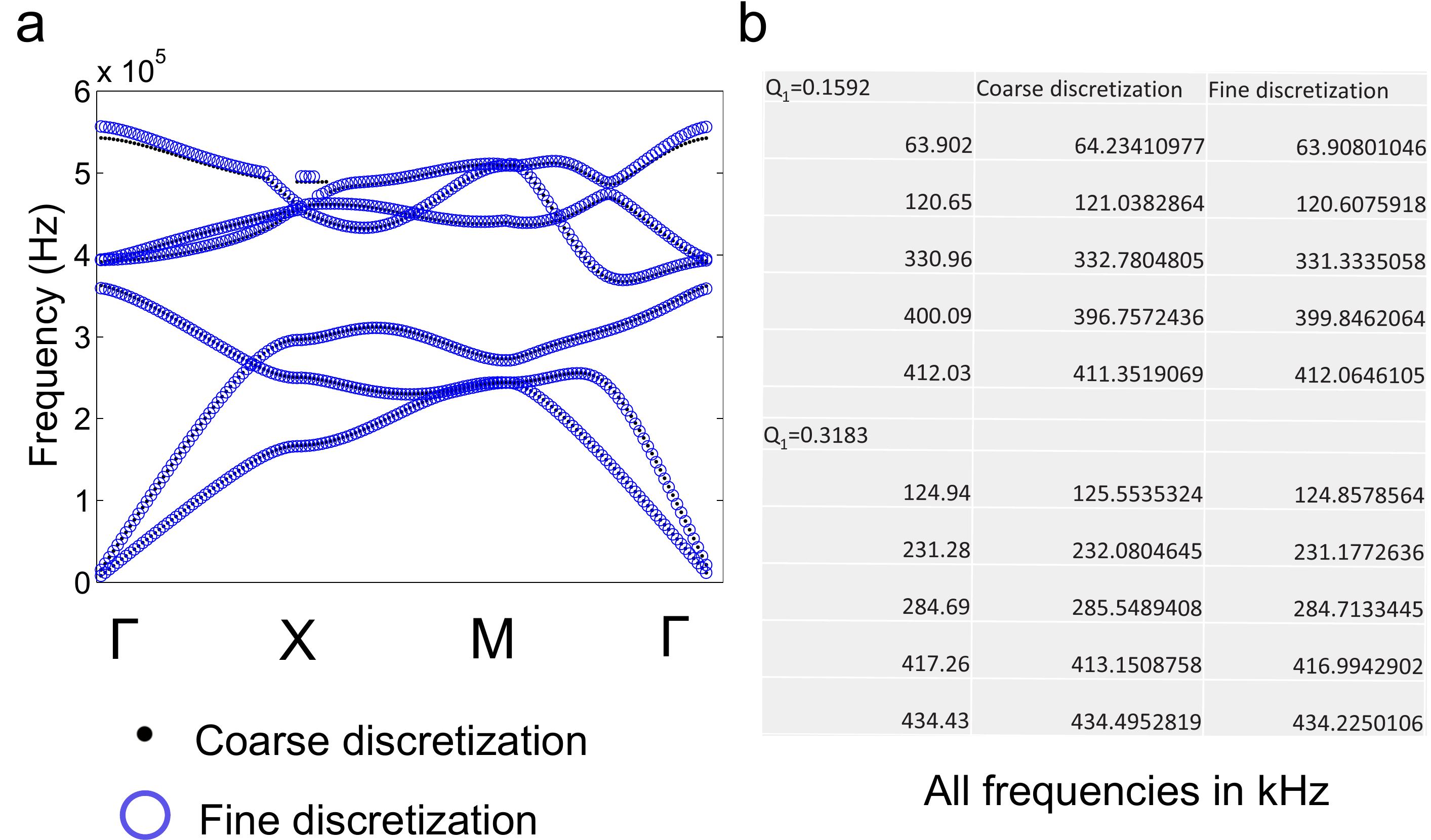}
\caption{a. Band-structure calculation for the coarse and the fine discretizations, b. Comparison with published results.}\label{fUnitCellSquareCircFibRes}
\end{figure}

Fig. (\ref{fUnitCellSquareCircFibRes}a) shows the results of the band-structure calculations for the coarse and the fine discretizations. The results are presented over the boundaries of the Irreducible Brillouin Zone ($\Gamma-X-M-\Gamma$). In these calculations we have taken $M=1$. This results in $2(2M+1)^2=18$ eigenvalues for each wave-number. In other words, once the matrices have been assembled for a chosen wave-vector, the eigenvalue problem results in the estimate of the first 18 frequencies. The method quickly converges to the solution for the first few frequencies. Depending upon the requirements higher number of terms in the expansion may be used to get better estimates for the higher frequencies. Although the method results in the first 18 frequencies for every wave-number, Fig. (\ref{fUnitCellSquareCircFibRes}a) shows the results for only the first 6 modes. The apparent discontinuities in the higher modes is not due to problems in the method but due to MATLAB's automatic ordering of the eigenvalues. The  results show that the effect of the higher level of discretization on the first few frequencies is negligible. 
The results for the 2 cases are virtually on top of each other for the first 5 modes and differ very slightly for the sixth mode. This indicates that for band-structure calculations over the first few modes it should be enough to employ rather rough discretizations. This is due to the fact that the Fourier functions for these low frequencies have small spatial variation and, therefore, it is reasonable to approximate them as being constant over larger triangles. The results are calculated at a total of 150 wave-number points and the average cpu time taken for the complete calculation is 21.46 seconds (coarse discretization).

Fig. (\ref{fUnitCellSquareCircFibRes}b) shows the comparison of the present calculations with those published earlier using exact expressions for the integrals \cite{nemat1975harmonic}. The comparisons are shown for the first five frequencies calculated at two wave-vectors ($Q_1=0.1592, Q_2=0; Q_1=0.3183, Q_2=0$). Due to a slight difference in the definition of the reciprocal basis these calculation points correspond to the normalized wavenumbers $Q=1,2$ used in the earlier study. The results in \cite{nemat1975harmonic} were presented as normalized frequencies and they have been converted to real frequencies for the purpose of comparison in this paper. It can be seen that even for the coarse discretization the results from the present calculations are very close to those calculated by employing exact expressions for the integrals. The advantage of the current method is a greatly increased flexibility of application of the method to arbitrarily complex geometries and unit cells. The calculation time, however, doesn't necessarily increase with the complexity of the unit cell. It depends only upon the level of discretization and the value of $M$. For instance, for the coarse discretization, the average cpu time taken for the complete calculation when $M=2$ increases to 86.37 seconds from 21.46 seconds when $M=1$. However it must be kept in mind that when $M=2$, we obtain the results for the first 50 frequencies as opposed to the first 18 when $M=1$.\\

\begin{table}[htp]
\caption {Comparison with exact integral calculations} \label{tCompare}
\begin{center}
\scalebox{.7}{
\begin{tabular}{ |c|c|c||c|c|c| }
  \hline
\multicolumn{6}{|c|}{$Q_1=0.1592$ (All frequencies in KHz)}\\ \hline
  \multicolumn{3}{|c|}{Square inclusion} & \multicolumn{3}{|c|}{Rectangular inclusion}\\
  \hline
  Exact integrals & Coarse mesh & Fine mesh & Exact integrals & Coarse mesh & Fine mesh\\ \hline
  63.96 & 63.82 & 63.95 & 63.92 & 63.87 & 63.81\\ \hline
  121.73 & 121.54 & 121.68 & 120.86 & 121.67 & 120.63\\ \hline
  323.42 & 321.22 & 323.28 & 328.31 & 323.65 & 328.01\\ \hline
  425.55 & 427.77 & 425.64 & 443.94 & 452.41 & 444.17\\ \hline
  429.16 & 432.24 & 429.14 & 502.63 & 512.76 & 502.95\\ \hline
  \hline
\multicolumn{6}{|c|}{$Q_1=0.3183$}\\ \hline
  \multicolumn{3}{|c|}{Square inclusion} & \multicolumn{3}{|c|}{Rectangular inclusion}\\
  \hline
  Exact integrals & Coarse mesh & Fine mesh & Exact integrals & Coarse mesh & Fine mesh\\ \hline
  124.31 & 123.84 & 124.37 & 120.86 & 120.69 & 120.99\\ \hline
  231.09 & 229.97 & 231.28 & 226.04 & 225.98 & 225.96\\ \hline
  285.25 & 285.58 & 285.10 &304.48 & 302.53 & 304.64\\ \hline
  439.48 & 440.79 & 439.82 & 452.08 & 463.05 & 452.54\\ \hline
  447.22 & 450.99 & 447.51 & 497.40 & 507.35 & 498.01\\ \hline
  \hline
\end{tabular}
}
\end{center}
\end{table}

Table (\ref{tCompare}) shows the comparison between the calculations from the current formulation and the results from \cite{nemat1975harmonic} for square and rectangular inclusions. The dimensions of the unit cell are the same as in the previous example. The dimensions of the inclusions are taken to correspond to the test cases used in Ref.\cite{nemat1975harmonic}.

\subsection{Example: hexagonal unit cell}

\begin{figure}[htp]
\centering
\includegraphics[scale=.23]{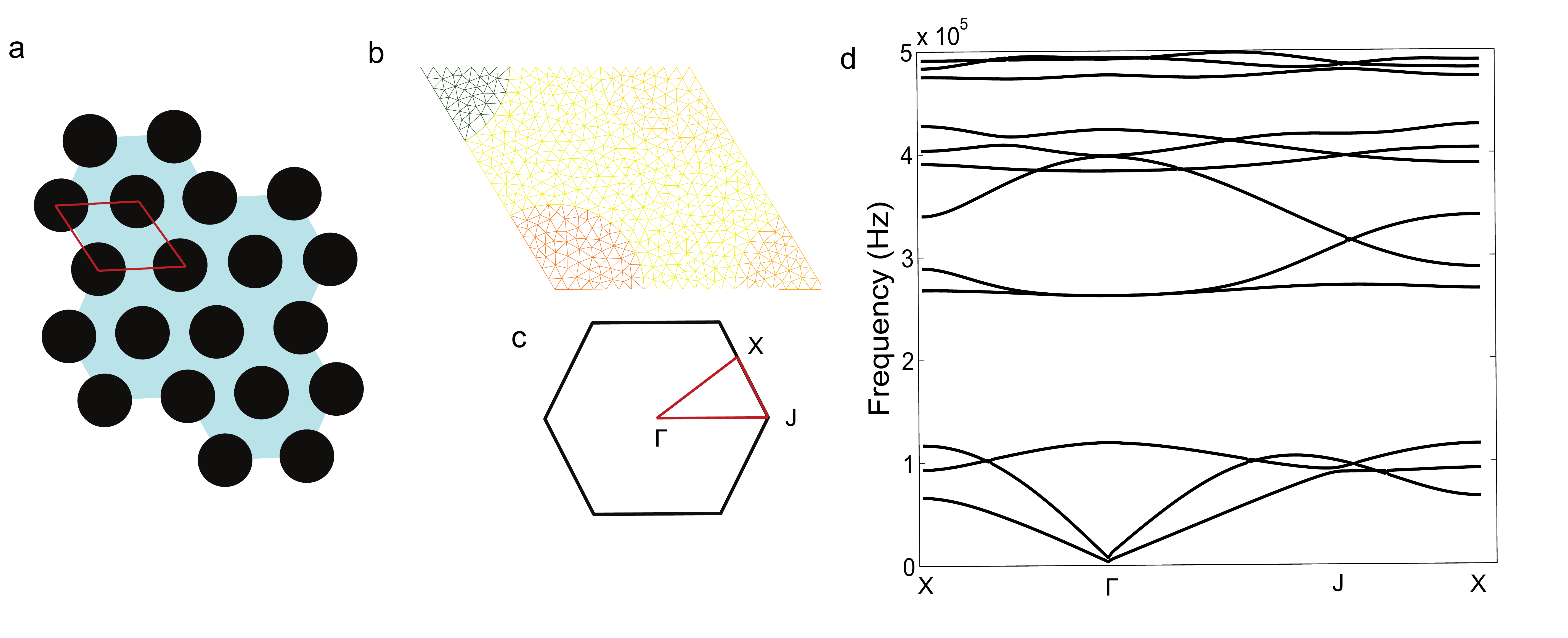}
\caption{a. Schematic of the 2-D periodic composite made from steel cylinders distributed in hexagonal packing in epoxy matrix, b. Discretization of the unit cell, c. Irreducible Brouillon Zone in the reciprocal lattice, d. Band-structure calculation results using the mixed variational formulation.}\label{fCompHexagonal}
\end{figure}

We now consider a hexagonal unit cell made up of steel cylinders ordered in an epoxy matrix (Fig. \ref{fCompHexagonal}a). The diameter of the steel cylinders is 4mm and the lattice constant is 6.023mm. The material properties are taken from \cite{vasseur2001experimental} and are reproduced here for reference

\begin{enumerate}
\item Steel: $C_{11}=264$ Gpa, $C_{44}=81$ Gpa, $\rho=7780$ kg/m$^3$
\item Epoxy: $C_{11}=7.54$ Gpa, $C_{44}=1.48$ Gpa, $\rho=1142$ kg/m$^3$
\end{enumerate}

Fig. (\ref{fCompHexagonal}b) shows the automatic discretization of the unit cell into 964 triangular elements. The band-structure is evaluated along the boundaries of the Irreducible Brouillon zone. This boundary is denoted by the path $X-\Gamma-J-X$ and is shown in Fig. (\ref{fCompHexagonal}c) in the reciprocal cell. For the band-structure calculations we use a total of 242 terms ($M=5$). This results in the simultaneous evaluation of the first 242 eigenvalues for each wavenumber point. The results in Fig. (\ref{fCompHexagonal}d), however, only show the first 12 eigenvalues. These results are in very good agreement with the PWE calculations shown by \cite{vasseur2001experimental} (Fig. 3 in that paper). We note the existence of the all-angle stop-band for waves traveling in the plane of the unit cell in the frequency ranges of 120-262 kHz and 427-473 kHz. The locations of the stop-bands and the general shape of the pass-bands are shown to match very well with the results of \cite{vasseur2001experimental}.

\section{Discussions and Conclusions}

In this paper we have presented an extension to the mixed-variational formulation which allows its use for the band-structure evaluation of 1-, 2-, and 3-D unit cells displaying arbitrary complexity in their Bravais structure and in the shape, size, number, and anisotropicity of their micro-constituents. Although the mixed-variational formulation has been shown to converge faster than displacement based variational formulations, it has never before been applied to solve for the band-structure of complex unit cells. We have elaborated upon the application of the method to 1-, and 2-D phononic composites (square and hexagonal unit cells) and verified the results with other established methods and from published literature. In particular, we have verified the results of the mixed-formulation in 1-D with the exact solution and in 2-D with the plane wave approximation. Although the results are presented for 1-, and 2-D unit cells, the formulation, as presented in this paper, can be easily applied to complex 3-dimensional unit cells.

\textbf{Acknowledgments}:
This research has been conducted at the Center of Excellence for Advanced Materials (CEAM) at the University of California, San Diego, under DARPA AFOSR Grants FA9550-09-1-0709 and RDECOM W91CRB-10-1-0006 to the University of California, San Diego.



\section{References}

\begin{thebibliography}{33}
\expandafter\ifx\csname natexlab\endcsname\relax\def\natexlab#1{#1}\fi
\expandafter\ifx\csname url\endcsname\relax
  \def\url#1{\texttt{#1}}\fi
\expandafter\ifx\csname urlprefix\endcsname\relax\def\urlprefix{URL }\fi

\bibitem[{Babuska and Osborn(1978)}]{babuska1978numerical}
Babuska, I., Osborn, J., 1978. Numerical treatment of eigenvalue problems for
  differential equations with discontinuous coefficients. Math. Comp 32, 991.

\bibitem[{Bilal and Hussein(2011)}]{bilal2011ultrawide}
Bilal, O.~R., Hussein, M.~I., 2011. Ultrawide phononic band gap for combined
  in-plane and out-of-plane waves. Physical Review E 84~(6), 065701.

\bibitem[{Bloch(1928)}]{bloch1928quantum}
Bloch, F., 1928. Quantum mechanics of electrons in crystal lattices. Z. Phys
  52, 555.

\bibitem[{Cervera et~al.(2001)Cervera, Sanchis, Sanchez-Perez, Martinez-Sala,
  Rubio, Meseguer, Lopez, Caballero, and
  S{\'a}nchez-Dehesa}]{cervera2001refractive}
Cervera, F., Sanchis, L., Sanchez-Perez, J., Martinez-Sala, R., Rubio, C.,
  Meseguer, F., Lopez, C., Caballero, D., S{\'a}nchez-Dehesa, J., 2001.
  Refractive acoustic devices for airborne sound. Physical review letters
  88~(2), 23902.

\bibitem[{Chan et~al.(1995)Chan, Yu, and Ho}]{chan1995order}
Chan, C., Yu, Q., Ho, K., 1995. Order-n spectral method for electromagnetic
  waves. Physical Review B 51~(23), 16635.

\bibitem[{Diaz et~al.(2005)Diaz, Haddow, and Ma}]{diaz2005design}
Diaz, A., Haddow, A., Ma, L., 2005. Design of band-gap grid structures.
  Structural and Multidisciplinary Optimization 29~(6), 418--431.

\bibitem[{Geuzaine and Remacle(2009)}]{geuzaine2009gmsh}
Geuzaine, C., Remacle, J.-F., 2009. Gmsh: A 3-d finite element mesh generator
  with built-in pre-and post-processing facilities. International Journal for
  Numerical Methods in Engineering 79~(11), 1309--1331.

\bibitem[{Goffaux and S{\'a}nchez-Dehesa(2003)}]{goffaux2003two}
Goffaux, C., S{\'a}nchez-Dehesa, J., 2003. Two-dimensional phononic crystals
  studied using a variational method: Application to lattices of locally
  resonant materials. Physical Review B 67~(14), 144301.

\bibitem[{Gorishnyy et~al.(2005)Gorishnyy, Ullal, Maldovan, Fytas, and
  Thomas}]{gorishnyy2005hypersonic}
Gorishnyy, T., Ullal, C., Maldovan, M., Fytas, G., Thomas, E., 2005. Hypersonic
  phononic crystals. Physical review letters 94~(11), 115501.

\bibitem[{Halkj{\ae}r et~al.(2006)Halkj{\ae}r, Sigmund, and
  Jensen}]{halkjaer2006maximizing}
Halkj{\ae}r, S., Sigmund, O., Jensen, J.~S., 2006. Maximizing band gaps in
  plate structures. Structural and Multidisciplinary Optimization 32~(4),
  263--275.

\bibitem[{Ho et~al.(1990)Ho, Chan, and Soukoulis}]{ho1990existence}
Ho, K., Chan, C., Soukoulis, C., 1990. Existence of a photonic gap in periodic
  dielectric structures. Physical Review Letters 65~(25), 3152.

\bibitem[{Hussein(2009)}]{hussein2009reduced}
Hussein, M., 2009. Reduced bloch mode expansion for periodic media band
  structure calculations. Proceedings of the Royal Society A: Mathematical,
  Physical and Engineering Science 465~(2109), 2825.

\bibitem[{Kafesaki and Economou(1999)}]{kafesaki1999multiple}
Kafesaki, M., Economou, E., 1999. Multiple-scattering theory for
  three-dimensional periodic acoustic composites. Physical Review B 60~(17),
  11993.

\bibitem[{Khelif et~al.(2003)Khelif, Choujaa, Djafari-Rouhani, Wilm,
  Ballandras, and Laude}]{khelif2003trapping}
Khelif, A., Choujaa, A., Djafari-Rouhani, B., Wilm, M., Ballandras, S., Laude,
  V., 2003. Trapping and guiding of acoustic waves by defect modes in a
  full-band-gap ultrasonic crystal. physical Review B 68~(21), 214301.

\bibitem[{Leung and Liu(1990)}]{leung1990full}
Leung, K., Liu, Y., 1990. Full vector wave calculation of photonic band
  structures in face-centered cubic dielectric media. Tech. rep., DTIC
  Document.

\bibitem[{Lin et~al.(2009)Lin, Huang, Sun, and Wu}]{lin2009gradient}
Lin, S., Huang, T., Sun, J., Wu, T., 2009. Gradient-index phononic crystals.
  Physical Review B 79~(9), 094302.

\bibitem[{Martinezsala et~al.(1995)Martinezsala, Sancho, Sanchez, G{\'o}mez,
  Llinares, and Meseguer}]{martinezsala1995sound}
Martinezsala, R., Sancho, J., Sanchez, J., G{\'o}mez, V., Llinares, J.,
  Meseguer, F., 1995. Sound-attenuation by sculpture. Nature 378~(6554),
  241--241.

\bibitem[{Minagawa and Nemat-Nasser(1976)}]{minagawa1976harmonic}
Minagawa, S., Nemat-Nasser, S., 1976. Harmonic waves in three-dimensional
  elastic composites. International Journal of Solids and Structures 12~(11),
  769.

\bibitem[{Minagawa et~al.(1981)Minagawa, Nemat-Nasser, and
  Yamada}]{minagawa1981finite}
Minagawa, S., Nemat-Nasser, S., Yamada, M., 1981. Finite element analysis of
  harmonic waves in layered and fibre-reinforced composites. International
  Journal for Numerical Methods in Engineering 17~(9), 1335--1353.

\bibitem[{Mohammadi et~al.(2008)Mohammadi, Eftekhar, Khelif, Hunt, and
  Adibi}]{mohammadi2008evidence}
Mohammadi, S., Eftekhar, A., Khelif, A., Hunt, W., Adibi, A., 2008. Evidence of
  large high frequency complete phononic band gaps in silicon phononic crystal
  plates. Applied Physics Letters 92~(22), 221905--221905.

\bibitem[{Nemat-Nasser(1972)}]{nemat1972harmonic}
Nemat-Nasser, S., 1972. Harmonic waves in layered composites. Journal of
  Applied Mechanics 39, 850.

\bibitem[{Nemat-Nasser et~al.(1975)Nemat-Nasser, Fu, and
  Minagawa}]{nemat1975harmonic}
Nemat-Nasser, S., Fu, F., Minagawa, S., 1975. Harmonic waves in one-, two-and
  three-dimensional composites: Bounds for eigenfrequencies. International
  Journal of Solids and Structures 11~(5), 617.

\bibitem[{Nemat-Nasser et~al.(2011)Nemat-Nasser, Willis, Srivastava, and
  Amirkhizi}]{nemat2011homogenization}
Nemat-Nasser, S., Willis, J., Srivastava, A., Amirkhizi, A., 2011.
  Homogenization of periodic elastic composites and locally resonant sonic
  materials. Physical Review B 83~(10), 104103.

\bibitem[{Reed et~al.(2003)Reed, Solja{\v{c}}i{\'c}, and
  Joannopoulos}]{reed2003reversed}
Reed, E., Solja{\v{c}}i{\'c}, M., Joannopoulos, J., 2003. Reversed doppler
  effect in photonic crystals. Physical review letters 91~(13), 133901.

\bibitem[{Rupp et~al.(2007)Rupp, Evgrafov, Maute, and Dunn}]{rupp2007design}
Rupp, C.~J., Evgrafov, A., Maute, K., Dunn, M.~L., 2007. Design of phononic
  materials/structures for surface wave devices using topology optimization.
  Structural and Multidisciplinary Optimization 34~(2), 111--121.

\bibitem[{Rytov(1956)}]{rytov1956acoustical}
Rytov, S., 1956. Acoustical properties of a thinly laminated medium. Soviet
  Physics-Acoustics 2, 68.

\bibitem[{Sigmund and Jensen(2003)}]{sigmund2003systematic}
Sigmund, O., Jensen, J.~S., 2003. Systematic design of phononic band--gap
  materials and structures by topology optimization. Philosophical Transactions
  of the Royal Society of London. Series A: Mathematical, Physical and
  Engineering Sciences 361~(1806), 1001--1019.

\bibitem[{Sukhovich et~al.(2008)Sukhovich, Jing, and
  Page}]{sukhovich2008negative}
Sukhovich, A., Jing, L., Page, J., 2008. Negative refraction and focusing of
  ultrasound in two-dimensional phononic crystals. Physical Review B 77~(1),
  014301.

\bibitem[{Vasseur et~al.(2001)Vasseur, Deymier, Chenni, Djafari-Rouhani,
  Dobrzynski, and Prevost}]{vasseur2001experimental}
Vasseur, J., Deymier, P., Chenni, B., Djafari-Rouhani, B., Dobrzynski, L.,
  Prevost, D., 2001. Experimental and theoretical evidence for the existence of
  absolute acoustic band gaps in two-dimensional solid phononic crystals.
  Physical Review Letters 86~(14), 3012--3015.

\bibitem[{White et~al.(1989)White, Wilkins, and Teter}]{white1989finite}
White, S.~R., Wilkins, J.~W., Teter, M.~P., 1989. Finite-element method for
  electronic structure. Physical Review B 39~(9), 5819.

\bibitem[{Yang et~al.(2002)Yang, Page, Liu, Cowan, Chan, and
  Sheng}]{yang2002ultrasound}
Yang, S., Page, J., Liu, Z., Cowan, M., Chan, C., Sheng, P., 2002. Ultrasound
  tunneling through 3d phononic crystals. Physical review letters 88~(10),
  104301.

\bibitem[{Yang et~al.(2004)Yang, Page, Liu, Cowan, Chan, and
  Sheng}]{yang2004focusing}
Yang, S., Page, J., Liu, Z., Cowan, M., Chan, C., Sheng, P., 2004. Focusing of
  sound in a 3d phononic crystal. Physical review letters 93~(2), 24301.

\bibitem[{Zhang and Satpathy(1990)}]{zhang1990electromagnetic}
Zhang, Z., Satpathy, S., 1990. Electromagnetic wave propagation in periodic
  structures: Bloch wave solution of maxwell's equations. Physical review
  letters 65~(21), 2650.

\end{thebibliography}
\end{document}